\documentclass[numberedappendix,12pt,preprint]{emulateapj} 
\slugcomment{Accepted for publication in ApJ - in press}  

\usepackage[]{amsmath}

\newcommand{\simgt}{\,\rlap{\lower 3.5 pt \hbox{$\mathchar \sim$}} \raise
1pt \hbox {$>$}\,}
\newcommand{\simlt}{\,\rlap{\lower 3.5 pt \hbox{$\mathchar \sim$}} \raise
1pt \hbox {$<$}\,}

\shorttitle{Quasar Color Variability}
\shortauthors{Schmidt et al. (2011)}

\usepackage{color}			      
\definecolor{midgray}{gray}{0.4}		
\definecolor{orange}{rgb}{1,0.5,0}    

\usepackage[pdftex,colorlinks=true,citecolor=midgray,linkcolor=midgray]{hyperref}        

\newcommand{\BE}{\begin{equation}}
\newcommand{\EE}{\end{equation}}
\newcommand{\BEA}{\begin{eqnarray}}
\newcommand{\EEA}{\end{eqnarray}}

\begin{document}


\title{The Color Variability of Quasars}


\author{Kasper B. Schmidt$^{1}$, Hans-Walter Rix$^{1}$, Joseph C. Shields$^{2}$, Matthias Knecht$^{1,3}$, David W. Hogg$^{1,4}$, Dan Maoz$^5$, Jo Bovy$^4$ }
\affil{$^{1}$ Max Planck Institut f\"ur Astronomie, K\"onigstuhl 17, D-69117 Heidelberg, Germany}
\affil{$^{2}$ Physics \& Astronomy Department, Ohio University, Athens, OH, 45701}
\affil{$^{3}$ Fakult\"at f\"ur Physik und Astronomie, Universit\"at Heidelberg, Albert-Ueberle-Str. 3-5, D-69120 Heidelberg, Germany}
\affil{$^{4}$ Center for Cosmology and Particle Physics, Department of Physics, New York University, 4 Washington Place, New York, NY 10003}
\affil{$^{5}$ School of Physics and Astronomy, Tel-Aviv University, Tel-Aviv 68878, Israel}

\email{kschmidt@mpia.de}




\begin{abstract}

We quantify quasar color-variability using an unprecedented
variability database -- $ugriz$ photometry of 9093 quasars
from SDSS Stripe 82, observed over 8 years at $\sim$60 epochs each.
We confirm previous reports that
quasars become bluer when brightening. We find a
redshift dependence of this blueing in a given set of bands (e.g. $g$ and $r$), but show that it is the result
of the flux contribution from less-variable or delayed emission lines in the different SDSS bands
at different redshifts. After correcting for this effect, quasar color-variability 
is remarkably uniform, and independent not only of redshift,
but also of quasar luminosity and black hole mass. The color variations of
individual quasars, as they vary in brightness on year timescales, are much more pronounced
than the ranges in color seen in samples of quasars across many orders of
magnitude in luminosity. This indicates distinct physical
mechanisms  behind quasar variability and the observed range of
quasar luminosities at a given black hole mass -- quasar variations cannot be explained by changes
in the mean accretion rate. We do find some dependence of the color variability on the
characteristics of the flux variations themselves, with fast,
low-amplitude, brightness variations producing more color variability. The observed
behavior could arise if quasar variability results from flares or
ephemeral hot spots in an accretion disc.

\end{abstract}



\section{Introduction}
\label{sec:intro}

Quasars, the brief phases of high accretion onto the massive black holes in the centers of large galaxies,
have proven to be one of the most versatile classes of astrophysical objects in the exploration of the distant Universe. Through large efforts and dedicated searches \cite[e.g.,][]{schmidt83,croom01,eyer02,richards02,richards04,atlee07,richards09,d'abrusco09,bovy11a} large samples of quasars are known today and have been explored in much detail to aid the understanding in fields as different as mass clustering on both large
and small scales \citep{croom05,croom09,shen07,shen09,ross09}, the understanding of the molecular
gas content in distant galaxies \citep{yun97,riechers07a,riechers07b}, and estimates of cosmological parameters and the dark energy equation of state \citep[e.g.,][]{scranton05,giannantonio08,xia09}.

Much effort has been put into understanding the nature of the quasars themselves and active galactic nuclei (AGN). 
Among the phenomena to be explained is the ubiquitous time-variability of the quasar emission.
Several physical processes have been invoked to explain the variability of the observed optical emission. 
Foremost are accretion disc instabilities
\citep[e.g.,][]{rees84,kawaguchi98,pereyra06}, but also large-scale changes in the amount of in-falling material may be important \citep[e.g.,][and references therein]{hopkins06}.
Various stochastic processes have also been suggested as possible causes with less success though.

There are different approaches on how to sort out which variability mechanisms are prevalent under what circumstance. The foremost diagnostic is the temporal behavior of flux variations:
Each of the mechanisms induces variability on different timescales, from weeks for changes on thermal timescales in the accretion disc, over months for superpositions of stochastic processes to several years for viscous changes in the large-scale structures of the accretion disc and for lens crossing times. These `physical' timescales of AGNs \citep[e.g.,][and references herein]{webb00,collier01} can be compared to the observed AGN variability timescales. The observed variability time-scales span the range from a few hours \citep{stalin04,gupta05}, possibly the result of processes in a jet \citep{kelly09}, to months and years where quasars are known to typically vary $\gtrsim10\%$ \citep[e.g.,][]{giveon99,collier01,vandenberk04,rengstorf04,sesar07,bramich08,wilhite08,bauer09,kelly09,kozlowski10}. However, it is a challenging task to disentangle the processes based on variability-timescales alone to get a clear view of the underlying processes.
Moreover, as AGN generally have power-law-shaped structure functions for the temporal variations, i.e., with no particular timescales, and because measurements of `characteristic variability timescales' are likely dominated, or at least influenced by window functions, interpretations of variability time-scales are always somewhat problematic. Nevertheless, there seems to be a general consensus in the community, that the most probable scenario for the majority of the observed variability behavior is changes in the accretion discs. 
 
The weekly to yearly variability of quasars (AGN) has been exploited for several independent purposes, e.g., for reverberation mapping \citep{peterson93,kaspi05,kaspi07} 
to estimate Eddington ratios and black hole masses \citep{peterson98,peterson04,kaspi00}, and for quasar identification \citep{scholz97,eyer02,geha03,sumi05}. Recently, new algorithms for identifying quasars via their variability have been established \cite[e.g.,][]{kelly09,schmidt10,palanque10,macleod10,macleod11,butler11,kim11}. These approaches provide an alternative to color selection alone for discovery of quasars \citep[e.g.,][]{richards02} in large-area multi-epoch surveys like the Panoramic Survey Telescope \& Rapid Response System \citep[Pan-STARRS;][]{kaiser02} and the future Large Synoptic Survey Telescope \cite[LSST;][]{ivezic08,LSST} where spectroscopic confirmation of the millions of quasar candidates to be found is unfeasible. 

Multi-wavelength AGN variability data to date provide clear albeit somewhat qualitative evidence that quasars tend to get bluer when they get brighter \citep[e.g.,][]{giveon99,trevese01,trevese02,geha03,vandenberk04,wilhite05,sakata11}. 
One practical caveat to such conclusions is that almost 
all photometric `brighter makes bluer' claims are based on fitting in flux versus color, without accounting for the color-magnitude error correlation in the modeling; such fitting may lead to spurious, or at least biased, results as we show in the present paper. To avoid these error correlations such analyses should be performed on flux-flux space. Here we develop a better unbiased fitting procedure to quantify whether brighter-makes-bluer on short time-scales \citep[see also][]{sakata11}.

In the present paper, we carry out a comprehensive study of `color variability' in quasars, i.e., we study how flux variability is linked to changes of the (observed) optical colors. We provide both a detailed empirical description of the observed variability, and work on linking it to the physics of the central engine by quantifying the correlation of color variability with the redshift of the quasars, their $M_\textrm{BH}$ and $L/L_\textrm{Edd}$, and to the temporal behavior of the flux variations.
While the temporal behavior has been studied extensively, there are no comprehensive studies 
of color variability in large quasar samples with many epochs of data, which offers great potential as a diagnostic of accretion disc physics.

The Sloan Digital Sky Survey \citep{stoughton02,gunn06} Stripe 82 provides an unprecedented data base for the present color variability study. It has a well defined quasar classification and several epochs (on average 60 epochs over 8 years) of precise photometry in five bands. 

The paper is structured as follows: after introducing the superb data set of Stripe 82 in Section~\ref{sec:data}, we describe the development of an unbiased fitting procedure in Section~\ref{sec:fit}. In Section~\ref{sec:results} we present the color variability results from applying this procedure to the Stripe 82 data. In Section~\ref{sec:disc} we discuss our findings and describe how the color variability depends on the light curve variability properties. Furthermore, we check for $L/L_\textrm{Edd}$ as well as $M_\textrm{BH}$ dependences, and compare our results to recent accretion disc models before we sum-up and conclude in Section~\ref{sec:conc}.  The central findings of the paper are reflected in Figures~\ref{fig:varmod}, \ref{fig:LM2Dhist}, and \ref{fig:AGzoom}.


\section{SDSS Stripe 82 Data}
\label{sec:data}

The Sloan Digital Sky Survey's (SDSS's) Stripe 82 (S82) is an equatorial stripe 2.5 degrees wide and about 120 degrees long which has been observed many times in the 5 SDSS bands over more than 8 years. 
The S82 data base hence provides an unprecedented collection of data for variability studies in general \citep[e.g.,][]{ivezic07,bramich08,sesar07,sesar10,bhatti10} and quasars in particular \citep{kelly09,schmidt10,macleod10,macleod11,butler11}. The analysis presented here is done on the $\sim$9,000 spectroscopically confirmed quasars in S82. They have been selected from the SDSS data archive\footnote{http://casjobs.sdss.org/CasJobs/default.aspx} as described in \cite{schmidt10}. 
Thus, we have a sample of quasars with on average 60 observations spread over a period of roughly 8 years. 
To obtain further information on each individual quasar, such as for instance estimates of the bolometric luminosity ($L_\textrm{bol}$) and black hole mass of the central engine ($M_\textrm{BH}$), we cross-matched our list of objects with the catalog of quasar properties presented in  \cite{shen10}. We found a total of 9093 matches which constitute the catalog we will use in the remainder of this paper, unless noted otherwise. This corresponds to basically the complete catalog of spectroscopically confirmed quasars in S82, hence, matching to the \cite{shen10} catalog did not cut down the sample much.

\section{Fitting Color Variability in Magnitude Space}
\label{sec:fit}

In principle optical \emph{color variability} of quasars, i.e., the tendency of changing color, generally becoming bluer when they brighten has been well established \citep{giveon99,vandenberk04,wilhite05}.
However, this color variability has been established and quantified by fitting data in color-magnitude space, for instance in $g$ versus ($g-r$) space, which suffers from co-variances between the color and the magnitude uncertainties that have not been accounted for in the past analyses. As we describe in Appendix~\ref{sec:fitting} we have found that this may lead to severe overestimates of the color variability, especially as the photometric errors are not negligible compared to the intrinsic variability amplitudes. To remedy these biases we fit the color variability in magnitude-magnitude space, and then `translate'  them into color-magnitude relations.

The S82 data presented in Section~\ref{sec:data} are unprecedented in their combination of time coverage, number of epochs, filter bands and sample size, which allows us to take the color variability analysis to the next level.
Since the quasar variability typically shows modest amplitude (a few tenths of a magnitude or less) it has been characterized in previous work by a linear relation in flux-flux space \citep[e.g][]{choloniewski81,winkler97,suganuma06,sakata10,sakata11}.
We make the \emph{ansatz} that the photometric measurements of each individual quasar can be represented by a linear relation in $gr$-space (and $ui$-space). 
The $gr$ and $ui$ spaces were chosen for high S/N and broad spectral range respectively (the $uz$-space being too uncertain due to $z$-band measurement errors). 
By calculating the Pearson correlation coefficient (PCC) for each individual quasar we verified that this approach is indeed sensible. Averaged over the full sample in $gr$-space the $\textrm{PCC}=0.8$. A PCC of 1 indicates a linear correlation with basically no scatter. Including an extra parameter in the linear relation to resemble the intrinsic scatter when determining $s_\textrm{gr}$ also shows that the scatter of the assumed linear relation is insignificant. Hence, the data does indeed resemble a linear relation quite closely.

When fitting such data a number of factors need to be taken into account: there are comparable errors along both axes, e.g., for $g$ and $r$ magnitudes, the errors vary widely among different data points, and there are `outliers' \citep[see][for physical, observational or data processing reasons]{schmidt10}. In order to take these factors properly into account we have used the linear fitting approach, including outlier pruning, laid out in \cite{hogg10} page 29ff. In practice we identify the set of relations that make the data likely outcomes
\BE\label{eqn:line}
r - \langle r\rangle   = s'_{gr}(g-\langle g\rangle) + b
\EE
by a Metropolis-Hastings Markov chain Monte Carlo (MCMC) approach. Thus, we are determining the color variability (slope) $s'_{gr}$ and the offset on the $r$-axis $b$ (moving each object to its mean $g$ and $r$ value to improve the determination of $b$) by sampling the parameter space via an MCMC chain. From simple algebraic manipulations of Equation~\ref{eqn:line} we have that
\BEA
r &=& s'_{gr}g + b' \\
g-r &=& -(s'_{gr}-1)\left(g-\langle g \rangle \right) + B 
\EEA
where $b' = \langle r \rangle - s'_{gr} \langle g \rangle + b$ and $B = - b + \left( \langle g \rangle - \langle r \rangle \right)$ are constants. These equations give the `transformation' of the fit between magnitude-magnitude space and color-magnitude space. 
If $(s'_{gr}-1) < 0$, i.e., if $s'_{gr} < 1$ the quasar gets bluer as it brightens. In the remainder of this paper we will use 
\BE
s_{gr}=(s'_{gr}-1)
\EE
as our definition of the color variability. In more general terms this corresponds to $s_{\lambda_1\lambda_2}\equiv \frac{\partial m_{\lambda_2}}{\partial m_{\lambda_1}}-1$ where the $\lambda$s refer to the photometric bands. This expression has the intuitive interpretation that $s_{gr}=0$ means no color variability, i.e., brightness variation at constant color, whereas $s_{gr}<0$ accounts for the bluer equals brighter trend (the most common trend in the data) and $s_{gr}>0$ implies objects that become redder when they brighten.

\begin{figure}
\epsscale{1.2} 
\plotone{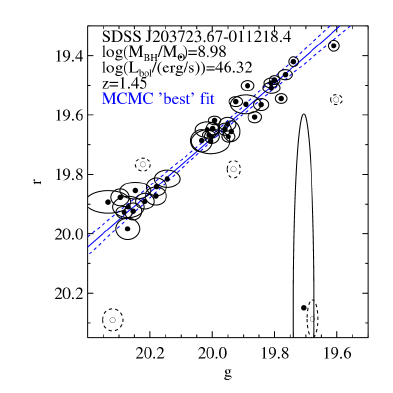}
\caption{Example of MCMC fitting for the $gr$ color variability, given a set of independently measured $g$ and $r$ light curve points and their uncertainties. These photometric data for the quasar SDSS J2037-0112 are shown as black circles with error ellipses indicating the photometric errors on the measurements. The solid blue line indicates the `best' MCMC fit (i.e., the parameters for which the data are most likely) of the linear relation in Equation~\ref{eqn:line} to the data.
The dashed blue lines show the 68\% confidence interval of the MCMC. 
The empty data points with dashed error ellipses have a posterior probability of being outliers to the relation of  $>50\%$, as described in Section~\ref{sec:fit}; see also Section~3 in \cite{hogg10}.
This fitting accounts for the independence of the $g$ and $r$ measurements, the widely varying error bars, and
`outliers'.
In the upper left corner we note the mass of the central black hole and the bolometric luminosity taken from the value-added quasar catalog presented in \cite{shen10} and the spectroscopic SDSS redshift of the object. In this manner the $gr$ and $ui$ color variability of each of the 9093 Stripe 82 quasars in the sample were determined.
}
\label{fig:MCMCfit}
\end{figure} 

In Figure~\ref{fig:MCMCfit} an example of such an MCMC fit to constrain the color variability is shown for the quasar SDSS J2037-0112 in $gr$ space. The filled circles represent the individual photometric epochs from S82, with the ellipses indicating the photometric errors in $g$ and $r$. 
The `best' fit from Equation~\ref{eqn:line} is shown as the blue solid line. The blue dashed lines show the 68\% confidence interval given by the 16th--84th inter-percentile range of the MCMC `cloud' of possible fits. The described fitting procedure also allows estimating the probability that a given data point is an outlier to the obtained relation, i.e., an estimate of the posterior probability that each individual observation `belongs' to the obtained relation \citep[see Section~3 in][]{hogg10}. Such outliers can be due to for instance weather, bad calibration, image defects etc. The data points represented by the open circles with the dashed error ellipses in Figure~\ref{fig:MCMCfit} have a posterior probability of being outliers to the shown MCMC fit which is larger than 50\%. 
On average 8\% and 19\% of the observed epochs were counted as outliers to the obtained relations when fitting in $gr$ and $ui$-space respectively.
Similar fits in $gr$ (and $ui$) space were performed for all 9093 quasars in the S82 sample, each resulting in estimates for the $gr$ and $ui$ color variability for each object. Note that the temporal ordering of the flux points plays no role in this analysis.

\section{Results}
\label{sec:results}

In the following subsections we will present the results from the investigation of the color variability, $s_{gr}$, $s_{ui}$, for the 9093 spectroscopic S82  quasars.

\subsection{Color Variability in $gr$}
\label{sec:gr}

\begin{figure*} 
\epsscale{1.1}
\plottwo{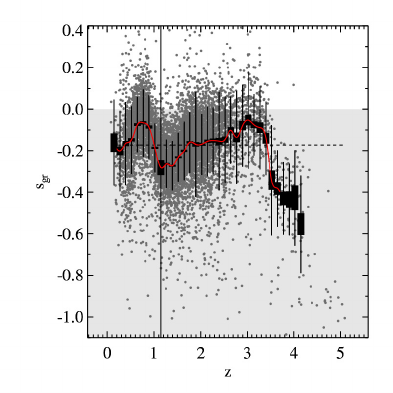}{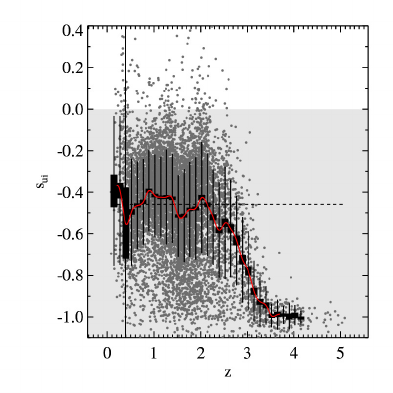}
\caption{The color variability, $s_{\lambda_1\lambda_2}$, in $gr$ (left) and $ui$-space (right) for the full sample of 9093 Stripe 82 quasars (dark gray dots), as a function of redshift. 
The black rectangles show the sample mean and its uncertainties in redshift bins of $\Delta z=0.125$, containing at least 5 data points.
The width ($\sigma$) of the color variability distribution is indicated by the thin error bars. 
The red curve gives the interpolated mean redshift trend, $\langle s_{gr} \rangle(z)$, and the black dashed line indicates the sample median color variability of -0.17 (-0.46) in $gr$ and $ui$. 
The shaded region indicates $s_{gr/ui}<0$, i.e., where bluer means brighter.
The complex redshift-dependence of color-variability is due to the influence of various emission lines (see discussion in Section~\ref{sec:gr} and \ref{sec:varmod}).}
\label{fig:zdep}
\end{figure*}

Figure~\ref{fig:zdep} shows the immediate result of the MCMC fitting procedure; the directly observed $gr$ color variability for the 9093 spectroscopically confirmed quasars from S82 as a function of their spectroscopic redshifts (left panel). It is clear that the vast majority of the quasars show color variability $s_{gr}<0$ (i.e., they get bluer when they brighten) represented by the shaded region. The right panel shows the analogous plot for the $ui$ color variability, which we consider further in Section~\ref{sec:ui}.

Figure~\ref{fig:zdep} shows that there is a very significant redshift dependence on the mean observed color variability, $\langle s_{gr/ui} \rangle (z)$.
%
Spectra \citep{wilhite05} and other information suggest that the $\langle s_{gr/ui}\rangle(z)$ behavior seen in Figure~\ref{fig:zdep} arises from a general trend of bluer continuum color in higher flux states modified by the redshift-dependent influence of emission lines in a given observed bandpass.  As we will show in detail below, such a description is consistent with trends in the S82 data.

\begin{figure*} 
\epsscale{1.1}
\plottwo{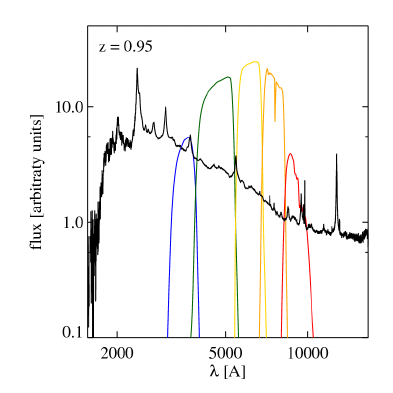}{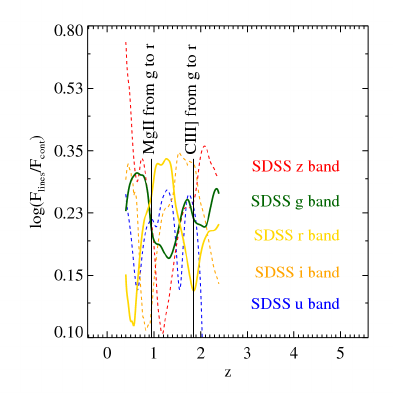}
\caption{The role of emission lines in the color variability. On the left, the \cite{vandenberk01} composite SDSS spectrum is shown with the 5 SDSS filter's response curves ($ugriz$ from left to right; arbitrarily scaled for visibility) as they would fall if the quasar was at $z=0.95$. In the right panel the expected ratio between the line flux $F_\textrm{line}$ and the continuum flux $F_\textrm{cont}$ for the same spectrum is shown with the SDSS $g$ (green) and $r$ (yellow) bands shown as solid curves and the $u$ (blue), $i$ (orange) and $z$ (red) bands shown as dashed curves. The redshift at which the MgII and CIII] lines move from the $g$ to the $r$ band ($z\sim0.95$ and $z\sim1.85$ respectively) has been indicated. 
This illustrates the redshift dependence of the color variability seen in Figure~\ref{fig:zdep} and described in the text. 
The redshift range in the right panel has been set equal to the one used in Figure~\ref{fig:zdep} to ease comparison.}
\label{fig:Fvsz}
\end{figure*}

The colors of quasars in general are known to have a pronounced redshift dependence resembling that seen in Figure~\ref{fig:zdep} because emission lines and the continuum affect static or single epoch colors \citep[e.g.,][]{richards01,wilhite05,wu10,meusinger11}.
For instance, the strong drop in $\langle s_{gr}\rangle(z)$ at $z\sim0.95$ in Figure~\ref{fig:zdep} corresponds exactly to the redshift where the MgII line moves from the $g$ to the $r$ band. This is illustrated in Figure~\ref{fig:Fvsz}, where the left panel shows the \cite{vandenberk01} composite quasar spectrum with the 5 SDSS bands as they would be positioned if the quasar was at $z=0.95$. The right panel shows the ratio between the emission line flux ($F_\textrm{line}$; the composite spectrum minus the estimated continuum flux) and the estimated continuum flux ($F_\textrm{cont}$; modeled as a simple power-law) as a function of redshift. The shift of the MgII line from the $g$ (green) to the $r$ (yellow) band is marked. Likewise the dips and bumps in $\langle s_{gr}\rangle(z)$ at $z\sim1.85, 2.8$ and $3.5$ in the left panel of Figure~\ref{fig:zdep} are attributable to the CIII], CIV and Ly$\alpha$ lines shifting between the $g$ and $r$ bands respectively. In the right panel of Figure~\ref{fig:Fvsz} only $z\lesssim2.4$ is shown since this is the region where a simple power-law approximation of the continuum is valid. For higher redshift the $g$ and $r$ bands move blue-ward of the Ly$\alpha$ line.

In order to isolate the continuum color variability for comparison with other quantities such as $L/L_\textrm{Edd}$ and $M_\textrm{BH}$ (Section~\ref{sec:LM}), we need to eliminate the source redshift dependence induced by the emission lines. This is done by `emission line correcting' the individual values of $s_{gr}$ and $s_{ui}$ by the quantity
\BE
\langle s \rangle - \langle s_{k} \rangle(z) \quad .
\EE
Here the first term is the mean color variability of -0.17 (-0.46) in $gr$ ($ui$) which is our stand-in for the emission line free color variability (dashed line(s) in Figure~\ref{fig:zdep}). The second term is the mean color variability for the sample capturing the mean redshift dependence of the sample as depicted by the red line in Figure~\ref{fig:zdep} for each of the individual quasars, $k$.

\subsection{Reproducing the Color Variability Redshift Dependence with Simple Variability Model}
\label{sec:varmod}

We now quantify to which extent a simple spectral variability model can reproduce the observed redshift trends in $\langle s_{gr/ui}\rangle(z)$. We do this by integrating a time-varying sequence of mock spectra created from the \cite{vandenberk01} composite quasar spectrum over the SDSS $g$ an $r$ filters as illustrated in the left panel of Figure~\ref{fig:Fvsz}.
After decomposing the \cite{vandenberk01} spectrum in a continuum and line component, by subtracting the estimated power-law continuum from \cite{vandenberk01}, we varied both the continuum and the lines to create a mock time sequence of spectra for which we obtained $g$ and $r$ light curves and then s$_{gr}$. 
By changing the slope of the continuum (with a pivot-point in the IR to ensure $s_{gr}<0$) and scaling the line response by a given amount, a sequence of spectra could be created to simulate a variable quasar. The line response was characterized by the ratio between the total integrated change in continuum flux and the total change in line flux over the modeled wavelength range 
\BE\label{eqn:alpha}
\alpha = \frac{\delta F_\textrm{line}}{\delta F_\textrm{cont}}
\EE
and could be set free (both lines and continuum can vary freely) or be fixed. Several setups for creating the sequence of variable spectra were inspected. Among those setups were fixed line contribution with changing continuum slope and both continuum and lines changing in various ways. 
For given $\alpha$ the emission lines are assumed to respond instantly to the continuum variation; 
i.e., in this simplistic approach we ignore any reverberation time-delay between the continuum and the lines.
An exploration of this effect to carry out reverberation mapping \citep[e.g.,][]{peterson04,kaspi05,kaspi07,chelouche11} using the broad-band light curves seems promising in light of Figure~~\ref{fig:zdep}, but is beyond the scope of this paper.
Details on the simple spectral variability models are given in Appendix~\ref{app:varmod}.  

\begin{figure}
\epsscale{1.2} 
\plotone{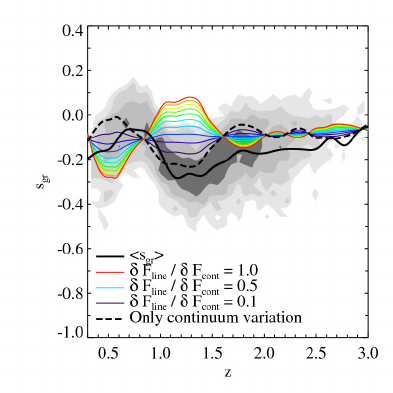}
\caption{Comparison of the observed redshift dependence of $s_{gr}$ with our simple spectral variability models. 
The mean $\langle s_{gr}\rangle(z)$ is shown by the solid black line (cf. Figure~\ref{fig:zdep}). 
The rainbow colored curves show the color variability models with fixed ratios $\alpha$ (see Equation~\ref{eqn:alpha})  between the line variability and continuum variability
from a ratio of 0.1 (purple) to a ratio of $\alpha=1.0$ (red). The variability model, where only the continuum varies ($\alpha=0$) is shown as the dashed black line, providing the best match to the observations.
Reverberation mapping time delays, \citep[$\sim 0.3...1 \;\textrm{year} \times\;(1+z)$ in quasars, e.g.,][]{kaspi07} are not taken into account.}
\label{fig:varmod}
\end{figure}

The predictions of the spectral variability models are shown in Figure~\ref{fig:varmod} together with the estimated values of $s_{gr}$, shaded regions, and the mean redshift dependence, $\langle s_{gr}\rangle(z)$, from the left panel of Figure~\ref{fig:zdep}. 
This Figure shows that $\langle s_{gr}\rangle(z)$ can be best matched if the (implicitly instantaneous) line response is very sub-linear: $\alpha=0.1$ (purple line in Figure~\ref{fig:varmod}) is a much better fit than the model with $\alpha=1$ (red line in Figure~\ref{fig:varmod}). 
It is seen that for emission lines that vary in lockstep with the continuum by $\alpha>25\%$ the redshift features in $\langle s_{gr}\rangle(z)$ are `inverted'. Actually, unresponsive line fluxes (i.e., $\alpha=0$), lead to the best match in this model context (black dashed curve in  Figure~\ref{fig:varmod}). Overall, Figure~\ref{fig:varmod} tells us that the redshift dependence of the $gr$ color variability is nicely reproduced by a simple spectral variability model where the continuum of the spectrum is hardened, i.e., its power-law slope is changed so brighter makes bluer, and the emission line fluxes in the two bands are (instantaneously) unresponsive.
We know from detailed reverberation studies that emission lines do respond \citep[on ~1/2 year timescales][]{kaspi05,kaspi07} for quasars at $z\sim1$. The explanation for $\alpha\approx 0$ may be that the lags in the lines are long enough to introduce a phase offset whereby the line is sometimes stronger, sometimes weaker than predicted from a tight correlation and in the net the correlation gets lost. 

Improved modeling, including a broken power-law continuum, explicit treatment of line reverberation and lack of variability at higher rest wavelengths as shown in \cite{vandenberk04} and \cite{wilhite05}, would be fruitful to carry out, but is beyond the scope of the present paper.

\subsection{Color Variability in $ui$}
\label{sec:ui}

\begin{figure}
\epsscale{1.2} 
\plotone{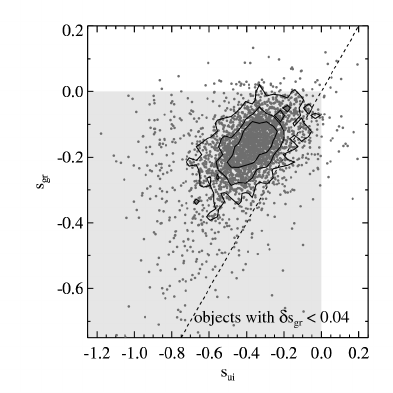}
\caption{The color-variability in $gr$ vs. $ui$ color variability space, after correcting for the line-induced redshift dependence (as described in Section~\ref{sec:gr}). The gray shaded region shows where $s<0$, i.e., where bluer means brighter. The dashed line indicates $s_{gr} = s_{ui}$ for reference: the $ui$ color variability is more pronounced on average than the $gr$ color variability. Only the 3111 objects of the sample with $\delta s_{gr}<0.04$ are shown. 
}
\label{fig:grVSui}
\end{figure}

The SDSS S82 data offer the opportunity to extend this analysis beyond the relatively short spectral range covered by $g$ and $r$, 4770\AA{} to 6231\AA{} in the observed frame. We do so by exploring the color variability in the $u$ versus $i$ magnitude-magnitude space, which covers a spectral range from 3543\AA{} to 7625\AA. We chose to use the $i$-band instead of $z$ because of the significantly smaller photometric uncertainties in the $i$-band. The fitting procedure was exactly analogous to the case of $gr$ color variability as described in Section~\ref{sec:fit}. The right panel of Figure~\ref{fig:zdep} shows the estimated $ui$ color variability for the S82 quasar sample. Despite the larger scatter in $s_{ui}$ at any given redshift we see similar features such as a distinct redshift dependence in $\langle s_{ui}\rangle(z)$ superimposed on quite dramatic overall color variability of -0.46. 
It is clear that the $ui$ color variability is more pronounced than the $gr$ color variability; $\langle s_{ui}\rangle(z)<\langle s_{gr}\rangle(z)$.
This holds true for the ensemble properties as well as for individual objects, as illustrated in Figure~\ref{fig:grVSui}, where we plot the emission line corrected (as described in Section~\ref{sec:gr}) color variability in $gr$ and $ui$. 
The fact that $s_{ui}<s_{gr}$ for almost all objects, implies that there is a relatively stronger blueing over the $ui$ spectral range than over the smaller $gr$ range. The same conclusion is reached when accounting for the difference in the wavelength baselines between $gr$ and $ui$ by normalizing $s_{gr/ui}$ with the corresponding wavelength ratio.

The simplistic model for the redshift dependence of the $gr$ color variability described in Section~\ref{sec:varmod} and shown in Figure~\ref{fig:varmod}, gives equally good results for $\langle s_{ui}\rangle(z)$.

\section{Discussion}
\label{sec:disc}

We now proceed to put the color variability into the context of other physical parameters that describe the quasar phase and the time dependence of variability, exploring to which physical processes color variability may be linked.

\subsection{Color Variability as a Function of Eddington Luminosity and Black Hole Mass}
\label{sec:LM}

All 9093 spectroscopically confirmed quasars have matches in the quasar catalog presented in \cite{shen10}, of which 99.9\% (9088) have an estimate of the bolometric luminosity and 84.1\% (7615) have an estimated black hole mass derived from MgII \citep[see][for details]{shen10}. This allows us to normalize the luminosity to the Eddington luminosity ($L_\textrm{Edd}$).

\begin{figure*}
\epsscale{1.10} 
\plottwo{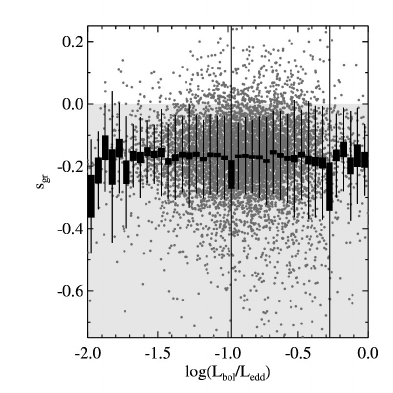}{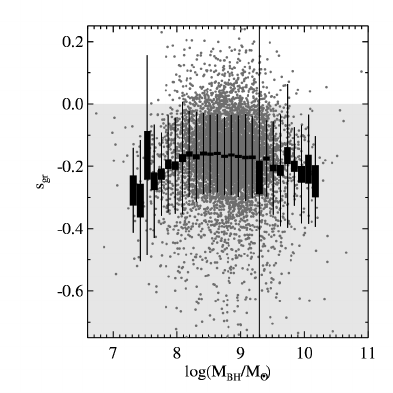}
\caption{The emission line corrected $gr$ color variability does not depend on $L/L_\textrm{Edd}$ (left) or on $M_\textrm{BH}/M_\odot$ (right). The black rectangles indicate the mean $s_{gr}$ and its error in each $L$ and $M_{BH}$ bin.
The thin error bars show the width of the $s_{gr}$ distribution cf. Figure~\ref{fig:zdep}.
No correlation between the color variability and $L/L_\textrm{Edd}$ and $M_\textrm{BH}/M_\odot$ is detected. This implies that the color variability is not a function of black hole mass or the overall accretion (Eddington) rate.
}
\label{fig:LandMrel}
\end{figure*}

\begin{figure}
\epsscale{1.1} 
\plotone{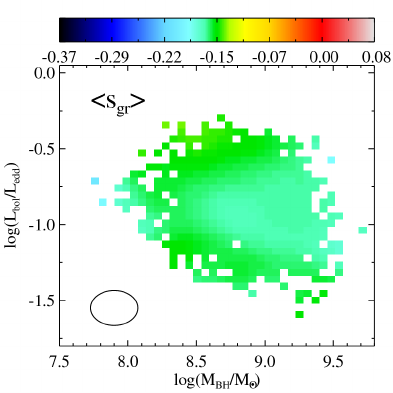}
\caption{The color-variability $s_{gr}$ as a function of $L/L_\textrm{Edd}$ and $M_\textrm{BH}/M_\odot$, as in Figure~\ref{fig:LandMrel}, again showing no significant trends. The color coding indicates the emission line corrected median $gr$ color variability of the objects in the bin. 
To reflect the uncertainty in luminosity and mass, the distribution has been smoothed by a 2D gaussian (represented by the ellipse in the bottom left corner) with full width at half maximum of $\textrm{FWHM} = [\textrm{FWHM}(M_\textrm{BH}),\textrm{FWHM}(L/L_\textrm{Edd})] = [0.35,0.24]$ corresponding to $[\sigma(M_\textrm{BH}),\sigma(L/L_\textrm{Edd})] = [0.15\textrm{dex},0.1\textrm{dex}]$.
}
\label{fig:LM2Dhist}
\end{figure}

If we now plot the emission line corrected (as described in Section~\ref{sec:gr}) $gr$ color variability against $L_\textrm{bol}$ and $M_\textrm{BH}$, as shown in Figure~\ref{fig:LandMrel}, it is evident that there is no detectable relation between the color variability $s_{gr}$ and the $L/L_\textrm{Edd}$ or $M_\textrm{BH}$. This is also illustrated in Figure~\ref{fig:LM2Dhist} where a 2D histogram of $L/L_\textrm{Edd}$ and $M_\textrm{BH}$, with the bins color coded according to the median $s_{gr}$, is shown: across the well sampled range in $L_\textrm{bol}$ and $M_\textrm{BH}$, the median $s_{gr}$ varies by no more than ~0.01 as a function of these two variable about its mean value of -0.17.
The 2D histogram has been smoothed by a 2D gaussian to reflect the uncertainty in luminosity and mass, with the full width at half maximum of the smoothing kernel (represented by the ellipse in the bottom left of Figure~\ref{fig:LM2Dhist}) $\textrm{FWHM} = [\textrm{FWHM}(M_\textrm{BH}),\textrm{FWHM}(L/L_\textrm{Edd})] = [0.35,0.24]$ corresponding to $[\sigma(M_\textrm{BH}),\sigma(L/L_\textrm{Edd})] = [0.15\textrm{dex},0.1\textrm{dex}]$.
Plots similar to the ones shown in Figure~\ref{fig:LandMrel} and \ref{fig:LM2Dhist} for the $ui$ color variability show no significant $L_\textrm{Edd}$ or $M_\textrm{BH}$ dependence, either. In Figure~\ref{fig:LandMrel} the full sample, i.e., all masses and redshifts are shown. Inspecting smaller sub-samples in $z$ (and $M_\textrm{BH}$) space does not change the picture. Hence, we find no correlation between the color variability in $gr$ (and $ui$) with  $L/L_\textrm{Edd}$ or $M_\textrm{BH}$.
More broadly, this seems to imply that the overall state of the quasar (characterized by $L/L_\textrm{Edd}$ and $M_\textrm{BH}$) plays no significant role in determining the color variability.

%
%

\subsection{The Color Variability as a Function of the Light Curve Variability Characteristics}
\label{sec:AG}

In \cite{schmidt10} we characterized the $r$-band variability of all the 9093 quasars through a `structure function' with an amplitude parameter $A$ and the light curve stochasticity, $\gamma$. The structure function variability of each individual quasar was modeled by a simple power-law
\BE
\textrm{SF}_\textrm{mod}(\Delta t_\textrm{obs} | A,\gamma) = 
  A \left( \frac{\Delta t_\textrm{obs}}{1 \textrm{yr}}\right)^\gamma
\EE
with $\Delta t_\textrm{obs}$ being the time between the observation of two individual photometric epochs and $\textrm{SF}_\textrm{mod}=\sqrt{ \langle (m(t_1) - m(t_2))^2 \rangle}$. The structure function of a periodically varying object or one varying like white noise will have a flat structure function and hence a small power-law exponent $\gamma$. 
Thus a large $\gamma$ indicates a secularly varying object or an object with a random walk like variability. The latter has been shown to describe quasar variability well in \cite{kelly09} and \cite{macleod11}.
The amplitude $A$ corresponds to the average variability on a 1 year timescale. 
In \cite{schmidt10} all calculations were done in the observed frame to mimic quasar identification with no prior information such as redshift. However, the spectroscopic redshift of each of the 9093 S82 quasars is known and the amplitude can be corrected for time-dilation. The rest-frame variability amplitude $A'$ is defined to be $A(1+z)^{\gamma}$ such that
\BE\label{eqn:powerlaw}
\textrm{SF}_\textrm{mod}(\Delta t_\textrm{rest} | A',\gamma) = 
  A' \left( \frac{\Delta t_\textrm{rest}}{1 \textrm{yr}}\right)^\gamma 
\EE
where $\Delta t_\textrm{rest}$ is now the difference between observations in the quasar rest-frame. All quoted $A'$s are estimated from the robust $r$-band measurements as described in \cite{schmidt10}. The variability amplitudes are independent of $z$ in agreement with \cite{giveon99} and the majority of the previous studies listed in their Table~1. 

In the following, however, the structure-function parameters A and $\gamma$ have been obtained
somewhat differently from \cite{schmidt10}. Rather than fitting the
structure function directly to the magnitude differences we fit a
Gaussian Process model \citep{rasmussen2006} defined by the structure
function to the magnitudes directly. This properly includes all of
the correlations between data points. This Gaussian Process model
consists of an $n$-dimensional Gaussian distribution (for $n$ epochs)
with a constant mean $m$ and $n$ by $n$ variance matrix $V$. The
elements of this variance matrix are given by
\begin{equation}
V_{ij} \equiv V(|t_i-t_j|) = V(\Delta t_{ij}) =\frac{1}{2} \left[ \textrm{SF}_{ij}^2(\infty) - \textrm{SF}_{ij}^2(\Delta t_{ij}) \right]
\end{equation}
for data points at epochs $t_i$ and $t_j$. Here the structure function $\textrm{SF}_{ij}$ is given by
\BE
\textrm{SF}_{ij} = \sqrt{ \langle (m(t_i) - m(t_j))^2 \rangle} \quad .
\EE
The photometric-uncertainty variances are added to the diagonal elements of $V$.
For the power-law structure function we cut off the power-law at 10 years such
that $\textrm{SF}(\infty)$ is finite. As all data span less than 10 years this
cut-off does not influence the fit. This type of fit is similar to the
Ornstein-Uhlenbeck process describing quasar variability as a damped
random walk \citep[e.g.,][]{kozlowski10,butler11,macleod11}. For
more details, see \cite{bovy11b}.

\begin{figure*}
\epsscale{1.10}  
\plottwo{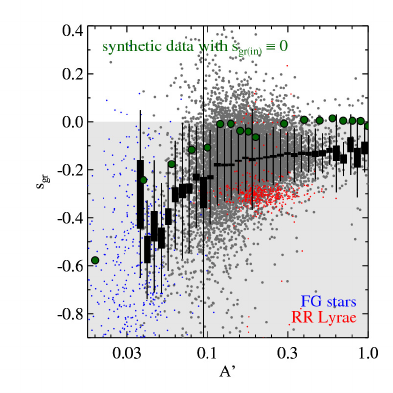}{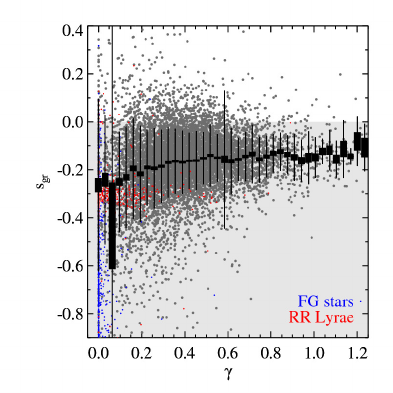}
\plottwo{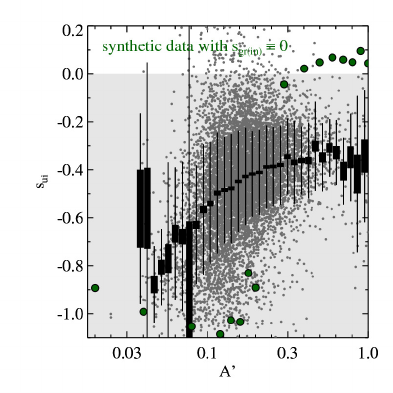}{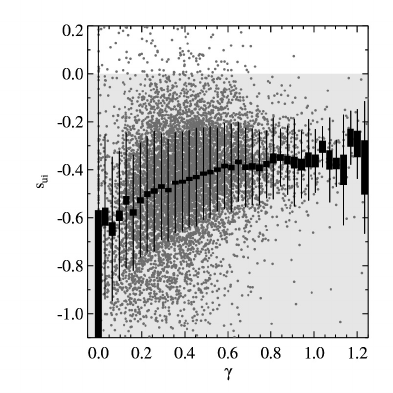}
\caption{Correlation between the light curve (temporal variability) structure function and color variability. The emission line corrected color variability in $gr$ (top) and $ui$ (bottom) are plotted against $A'$ (left) and $\gamma$ (right) from \cite{schmidt10}. $A'$ indicates the mean level of variability within one year (rest-frame), and $\gamma$, the power law exponent in the structure function, indicates how random (low $\gamma$) or secular (high $\gamma$) the light curve variations are.
Gray dots (data), black rectangles and error bars are analogous to Figure~\ref{fig:LandMrel}.
In the $gr$ plots (top) 500 non-varying FG stars and the 483 RR Lyrae from \cite{sesar10} are shown on top of the 9093 S82 quasars (gray dots) as blue and red points respectively. 
In the left column the recovered average $s_{gr}$ ($s_{ui}$) for 50 FG stars is shown as green filled circles, where synthetic brightness variations with $s_{gr}(\textrm{in})\equiv 0$ and different variability amplitudes $A'_\textrm{sim}$ had been created. As described in the text this illustrates that only $gr$ ($ui$) trends for $A'\gtrsim0.1$ ($A'\gtrsim0.25$) can and should be trusted. 
It shows that objects with large $A'$, i.e., with large variability amplitudes, have a color variability close(r) to 0, i.e., less blueing when brightening, than do objects with small $A'$. The trends in the two right hand plots are dominated by low $A'$ objects and is therefore not trustworthy.}
\label{fig:AG}
\end{figure*}

We can now look at the emission line corrected $s_{gr}$ and $s_{ui}$ as a function of $A'$ and $\gamma$ for all quasars. Figure~\ref{fig:AG} shows that $s_{gr/ui}$ seemingly vary both with $A'$ and with $\gamma$.
However, the limit of little variability (small $A'$) requires particular care, both because outliers play a bigger role and because $A'$ and $\gamma$  starts to be degenerate \citep{schmidt10}. We estimated the $gr$ color variability of 500 color selected non-varying FG-stars (see \cite{schmidt10} for further details) and of 483 S82 RR Lyrae stars from  \cite{sesar10}. These are over-plotted in the top panel of Figure~\ref{fig:AG} as the blue and red points respectively. As expected the RR Lyrae have a well defined color variability, whereas the inferred color variability of the non-varying FG-stars span a much wider range of $s_{gr}$. Interestingly,  the majority of the non-varying FG-stars have color variability estimates of $s_{gr}<0$ like the quasars and the RR Lyrae stars. This seems to be caused by the outliers in $g$ being relatively larger than the outliers in $r$, hence affecting the initial guess of the MCMC in a bluer-brighter direction. 
In the case of the RR Lyrae the well defined mean color variability in $gr$ is expected as RR Lyrae change their effective temperature and luminosity during their pulsation. By creating a sequence of black body spectra with temperatures from 6200K to 7200K, estimating the flux received in the $g$ and $r$ bands for each spectrum, and using that as a simple model for a variable RR Lyrae star, a color variability of $s_{gr} \sim -0.23$ is obtained, in very good agreement with the observations (Figure~\ref{fig:AG}, top panel, red dots).
Thus, in general the $s_{gr}$ for the FG and RR Lyrae stars look as expected. 

\begin{figure*}
\epsscale{1.10} 
\plottwo{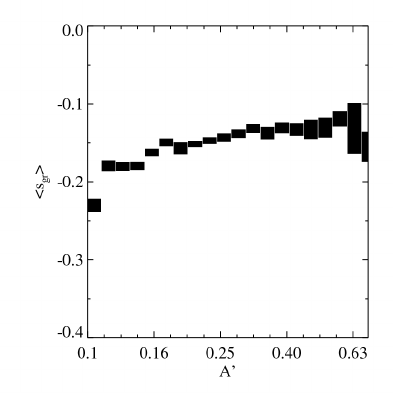}{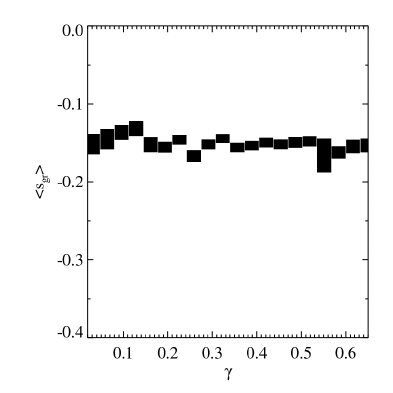}
\plottwo{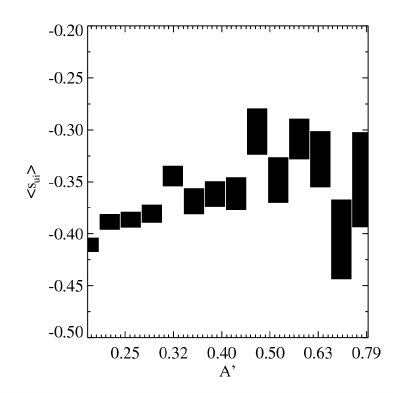}{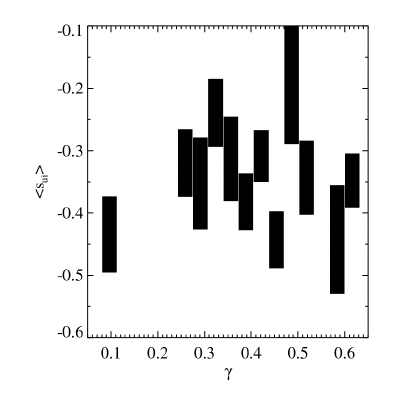}
\caption{The estimated correlations between mean color variability and structure function parameters. The four panels show the portion of parameter space in Figure~\ref{fig:AG} that is well populated by the sample. The rectangles in the left hand plots correspond to the ones in Figure~\ref{fig:AG}. In the right hand plots the rectangles are only estimated from objects with $A'>0.1 \; (0.25)$. A clear trend is seen, that the color variability of objects with high variability amplitude (left panel) are closer to 0 than objects with small variability amplitude. This indicates that strongly varying quasars get less blue as they brighten on average, than do moderate varying quasars. The color variability is independent of the power law index $\gamma$.}
\label{fig:AGzoom}
\end{figure*} 

A more direct way to estimate the fidelity of color variability estimates at low $A'$ values is to recover $s_{gr}$ estimates for objects of known (simulated) color variability. We induce such simulated variability into the  500 FG-stars by generating data of a certain 1 year amplitude ($A'$) from the original FG-star $g$ and $r$ photometry. We do that by generating new $u,g,r$ and $i$ magnitudes for the individual epochs $j$ via the expression
\BE
p_{j,\textrm{sim}} = p_j + A'_\textrm{sim}\frac{\Delta\textrm{MJD}_j}{365.25} \quad ,
\EE
by construction a data set with $s = 0$.
Here $p$ represents the photometric measurements in a given band, $j$ runs over the individual epochs, and $\Delta\textrm{MJD}_j$ refers to the observation time of the $j$th epoch with respect to the first observation. 
In this way many of the aspects of the real data (i.e., the outliers and realistic photometric errors) are included in the simulated data.
In the two left panels of Figure~\ref{fig:AG} this recovered mean average (over 50 randomly chosen FG stars) color variability is shown for a sequence of variability amplitudes, $A'_\textrm{sim}$, as large filled green circles. This shows that the recovered $gr$ ($ui$) color variability has some systematic errors below $A'\lesssim 0.1$ ($A'\lesssim 0.25$). 
When ignoring quasars with variability amplitudes smaller than  0.1 (0.25) the correlation between $s_{gr}$ and $s_{ui}$ and $A'$ is still present and significant. 
The trustworthy part of the relations, $\langle s_{gr/ui}\rangle(A,\gamma)$, in Figure~\ref{fig:AG} is shown in Figure~\ref{fig:AGzoom}. Again the black rectangles represent the uncertainty on the estimated mean color variability. The left hand plots correspond directly to Figure~\ref{fig:AG}, whereas for $\langle s_{gr/ui}\rangle(\gamma)$ the black rectangles are estimated only from objects with $A'> 0.1$ ($A'> 0.25$).
Figure~\ref{fig:AGzoom} clearly illustrates the trend that objects with larger variability amplitude have a smaller color variability (meaning less blueing when brightening) than for low $A'$. On the other hand the color variability is independent of $\gamma$.

As mentioned \cite{vandenberk04} and \cite{wilhite05} showed that there is a lack of variability at high rest wavelengths. Furthermore, it is known that quasar variability is anti-correlated with luminosity \cite[e.g.,][]{hook94,cristiani96,vandenberk04}.
This might lead to the suspicion that the trend presented in Figure~\ref{fig:AG} and \ref{fig:AGzoom} is nothing more than a redshift effect. If this was the case the relation should be due mainly to low-$z$ objects, since the most variable quasars are supposedly low luminosity quasars, i.e., necessarily only observed at low $z$, and should therefore disappear at high redshift. Estimating the relation between $s_{gr}$ and $A'$ in various redshift bins (also split in luminosity) shows that the trend is equally strong for all redshifts and all luminosities. Hence, the presented relation appears to be of a physical origin and not merely a redshift effect.
%

\subsection{Color Variability and Changes in the Mean Accretion Rate}
\label{sec:slopes}

In this Section we carry out a cursory exploration as to the physical origin of the observed color-variability. 

\subsubsection{Color Variability of Individual Quasars vs. the Color Distribution of Quasar Ensembles}

The colors of quasars at a given redshift are known to depend only weakly on their mean accretion luminosity or accretion rate \citep{davis07}, 
while we find that individual quasars become considerably bluer when they brighten on year time-scales.  This suggests different physical mechanism creating the accretion luminosity range in ensembles and the luminosity variations in individual quasars.

\begin{figure*}
\epsscale{1.10} 
\plottwo{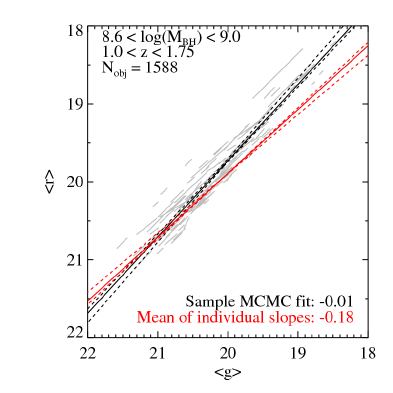}{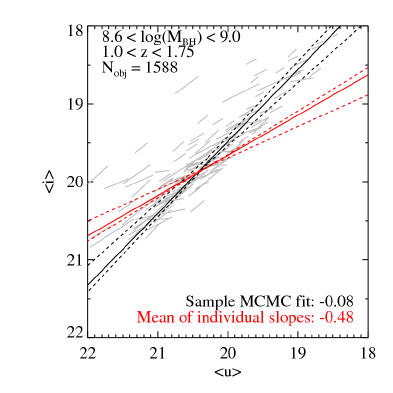}
\caption{The color-variability in individual quasars (red) vs. the color-luminosity relation in the ensemble of quasars (black), drawing on the sub-sample of 33rd--66th percentile of masses and the 25th--75th percentile of redshifts. The mass and redshift ranges are shown in each panel. 
Each individual quasar is indicated by a short solid gray line corresponding to the fitted and emission line corrected $gr$ ($ui$) relation, with a length reflecting the standard deviation of the $g$ ($u$) light curve. Only every 10th object of the sub-sample is shown to keep the individual objects visible. 
The red solid lines show the average trend within this sub-sample of $s_{gr}\sim-0.18$ and $s_{ui}\sim-0.48$ with the dashed red lines indicating the bootstrapping uncertainty.
The black lines show an MCMC fit to the time-averaged fluxes for the sample, i.e., the fit to $(\langle g \rangle, \langle r \rangle) \pm (\textrm{stdev}(g),\textrm{stdev}(r))$ for each quasar corresponding to $s_{gr}\sim-0.01$  and $s_{ui}\sim-0.08$. The dashed black lines show the 68\% confidence interval of the MCMC fit. 
In other words, the red line is the ensemble mean color variability while the black line is a fit to the ensemble of time-averaged mean magnitudes. 
Hence, it is clear that the average color variability of the individual quasars deviate significantly from the time averaged sample color variability. 
}
\label{fig:rVSg}
\end{figure*}

In Figure~\ref{fig:rVSg} the emission line corrected color variability of a sub-sample of the S82 quasars is shown in $gr$ and $ui$-space. 
This sub-sample represents the `average' quasars, i.e., the combined sample of the 33rd--66th percentile of masses and the 25th--75th percentile of redshifts for the quasar sample.
The color variability of each individual quasar is depicted as a short solid gray line showing $s'_{gr}$ ($s'_{ui}$) from Equation~\ref{eqn:line} for each quasar centered on $[\langle g \rangle,\langle r \rangle]$ ($[\langle u \rangle,\langle i \rangle]$) for that particular quasar. Only every 10th object of the sub-sample is actually shown to keep the individual gray lines visible. The length of the lines resembles the change in the photometric $g$-band ($u$-band) data of the quasar. 
The Figure compares the average $s'_{gr}$ ($s'_{ui}$) of all the individual quasars in the sub-sample (red solid line) with a fit to the time-averaged color distribution of the sub-sample (black solid line) where each data point corresponds to $[\langle g\rangle,\langle r\rangle]_k$ ($[\langle u\rangle,\langle i\rangle]_k$) with $k$ counting the quasars. Figure~\ref{fig:rVSg} reveals that indeed the mean color variability for individual quasars is much more pronounced than the equivalent quantity for the ensemble $s_\textrm{sample}=\frac{d\langle m_r\rangle}{d\langle m_g\rangle}$. 
For the given sub-sample $s_{gr}\sim-0.18$ and $s_{ui}\sim -0.48$ on average (as opposed to $s_{gr}\sim-0.17$ and $s_{ui}\sim -0.46$ for the full sample) compared to  $s_{gr}=-0.01$ and $s_{ui}\sim-0.08$ for the corresponding time-averaged sub-sample color distribution. 
The difference is highly significant in both cases, with the $ui$ color variability difference formally larger, because of the broader spectral range. The exact same trends are found for plots containing the full quasar sample.

This result shows that (temporal) color variability of individual quasars is considerably stronger than the color range of ensembles of quasars at similar redshifts and with similar black hole masses, that presumably differ in $L/L_\textrm{Edd}$. 

\subsubsection{Color Variability vs. Accretion Disc Models}
\label{sec:models}

Explaining quasar spectral energy distributions, and in particular the optical/UV continua through steady-state accretion disc models has an established history \citep[e.g.,][]{shakura73,bonning07,davis07}.
However, comparing the observed color variability of large samples of quasars with the predicted colors of model sequences of varying accretion rate has not been done yet. 
Such a comparison could tell us whether it is sensible to think of the quasar variability on scales of years as changes in the mean accretion rate.
The superb S82 data enables us to perform such a comparison, by comparing the observed color and color variability of the S82 quasars with sequences of accretion disc models presented in \cite{davis07}.

\cite{davis07} presented three different  thin accretion disc models that describe the spectral slope of quasars as a function of $L_\textrm{bol}/L_\textrm{Edd}$ and $M_\textrm{BH}$. We took these three models and worked out predictions for the observed $g$ and $r$ band for models of a given $M_\textrm{BH}$ but varying accretion rates.
The three models presented in \cite{davis07} and the color we adopt for their graphical representation are:
\begin{itemize}
\item[1)] A relativistic model of accretion onto a Schwarzschild black hole with a spin parameter of 0. The emission is based on Non-LTE atmosphere calculations
 (\textcolor{green}{green}).   
\item[2)] A relativistic model of accretion onto a Schwarzschild black hole with a spin parameter of 0. The disc is emitting as a black body 
(\textcolor{red}{red}).    
\item[3)] A Model of accretion onto a spinning black hole (spin parameter of 0.9) with emission based on Non-LTE atmosphere calculations 
(\textcolor{orange}{orange}).    
\end{itemize}
For further details on the models we refer to \cite{davis07}.

\begin{figure*}
\epsscale{1.10} 
\plottwo{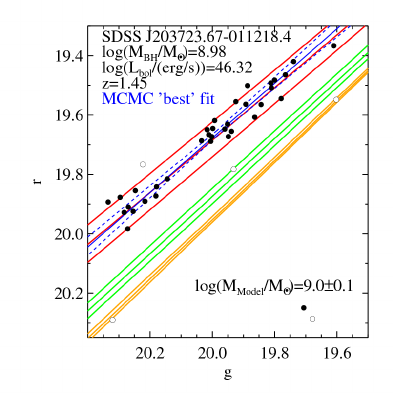}{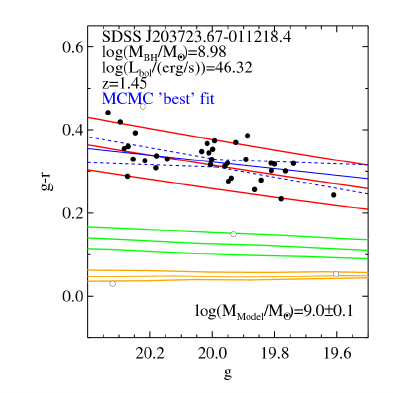}
\caption{Comparison of the observed color variability with sequences of steady-state accretion disc models \citep{davis07} of different accretion rates. This comparison is illustrated using the quasar SDSS J2037-0112, also shown in Figure~\ref{fig:MCMCfit}, shown in $r$ vs. $g$ (left) and $g-r$ vs. $g$ (right) space. The black symbols are the individual photometric measurements. The blue solid line shows the MCMC fit in $gr$-space to these measurements and the blue dashed lines indicate the 68\% confidence interval of that fit.
The open symbols denote likely outliers (see also Figure~\ref{fig:MCMCfit}).  The  \cite{davis07} models described in Section~\ref{sec:models} are shown in green (model 1), red (model 2) and orange (model 3) with $L/L_\textrm{Edd}$ changing along the lines. Each set of 3 lines denotes models for differing black hole masses near the value determined from the MgII line width \citep{shen10}. 
From top to bottom each set of lines (models) correspond to 9.1, 9.0 and 8.9 $\log(M_\textrm{model}/M_\odot)$.
%
For this object the match with model 2 is good; see Figure~\ref{fig:goodness} for an ensemble comparison.}
\label{fig:3modVSobj}
\end{figure*} 

We can then compare the models to the data in two respects: do they predict the right color (which has been done before) and do they predict the right change of color with changing accretion rate or luminosity?
In Figure~\ref{fig:3modVSobj} the object from Figure~\ref{fig:MCMCfit} is shown in $g$-$r$-$(g-r)$ space (without error ellipses) together with its best fit color variability (blue solid line). The three accretion disc models are shown as solid lines in bundles of three, where each of the three lines corresponds to a different black hole mass, as noted in the bottom right corner of each panel. In this particular case model 2) matches the data well both in color and in the change of color with changing luminosity. However, such a good match is not representative for the ensemble. We quantify this for the whole sample by estimating the `goodness' of the models as:
\BEA
\textrm{\textbf{D}}_{y} &=& \left(\sum_j^{N_k}\frac{1}{\delta y_j}\right)^{-1} \sum_j^{N_k} \frac{ y_j - y(x)_{j,\textrm{model}} }{\delta y_j} \label{eqn:D}\\
\Delta s &=& s_k - s_{k,\textrm{model}} \label{eqn:delta}
\EEA
where $y=(g-r)$ and $x=g$. The index $j$ runs over the $N_k$ epochs for each individual quasar $k$. The model prediction is the color at a given luminosity (or accretion rate), for a fixed black hole mass $y(x)_{j,\textrm{model}}$. 
The photometric error on the color for the $j$th measurement  is denoted as $\delta y_j$. 
Since the error on the $M_\textrm{BH}$ estimates based on MgII  \citep{shen10} is $\sim0.4$dex \citep{vestergaard06,deRosa11}, we have chosen to show three values for $M_\textrm{BH}$, leading to three model prediction lines, for each model in Figure~\ref{fig:3modVSobj}.

The `goodness' parameters $\textrm{\textbf{D}}_{y}$ and $\Delta s$ defined in equations~\ref{eqn:D} and \ref{eqn:delta} therefore describe how well the observed values of color and color variability are predicted by the \cite{davis07} models. $\textrm{\textbf{D}}_{y}$ can be seen as the standard $\chi^2$ measure of comparison between model and data \emph{before} squaring, i.e., it estimates the difference between the model color and the observed color averaged over all epochs for each quasar. The $\Delta s$ is simply the difference between model color variability and observed color variability of each quasar. 

\begin{figure}
\epsscale{1.1} 
\plotone{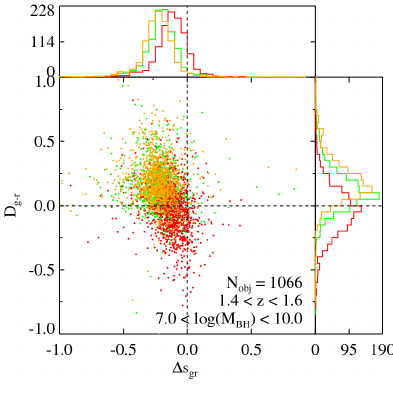}
\caption{Comparison of the observed color variability to steady-state accretion disc models with changing $L/L_\textrm{Edd}$ 
Each individual point reflects the data model discrepancy in the mean color $\textrm{\textbf{D}}_\textrm{g-r}$ (Equation~\ref{eqn:D}) and $\Delta s_{gr}$ (Equation~\ref{eqn:delta}) for one individual quasar in the parameter space illustrated in Figure~\ref{fig:3modVSobj}. The plot shows the sub-sample of 1066 quasars with $z=1.4$--$1.6$. The color of the points correspond to model 1) (green), 2) (red) and 3) (orange) from \cite{davis07} converted into $g$ and $r$ magnitudes. The dashed lines show perfect agreement between model and data in color and color variability, respectively. 
The projections of the distributions are shown as histograms along the axes.
While the different sets of models can reproduce (by design) the mean quasar colors (i.e., $\textrm{\textbf{D}}_\textrm{g-r}$), the observed color variability is far too strong to be interpreted as changes in accretion rate in a steady-state model context.
}
\label{fig:goodness}
\end{figure}

Figure~\ref{fig:goodness} summarizes the model data comparison for a subset of quasars with $z\sim1.5$: the quantities from equations~\ref{eqn:D} and \ref{eqn:delta}. 
All models predict a color variability -- as a function of changes in the mean accretion rate -- that is weaker than observed.
On average model 2) shown in red matches the observed $gr$ color variability the best. Furthermore, the $\textrm{\textbf{D}}_{g-r}$ values indicate that the $g-r$ color is on average overestimated by model 1) and 3), whereas the distribution of model 2) has a mean very close to the dashed perfect agreement line. 
Creating similar plots for other mass and redshift ranges as well as for the results in $ui$-space show the exact same trends. Thus, of the three \cite{davis07} accretion disc models considered here, model 2) matches the observed color and
the obtained $gr$ and $ui$ color variability the best.

\section{Conclusion}
\label{sec:conc}
In the present study we determined and analyzed the color-variability of 9093 spectroscopically confirmed quasars from SDSS Stripe 82, to understand to which extent and why quasars get bluer (redder) if they brighten (dim), by fitting linear relations between the SDSS $g$ and $r$ bands as well as between the $u$ and $i$ bands in magnitude-magnitude space.
The connection of various quasar properties to the color variability were inspected before the results were compared to models of accretion disks with varying accretion rates from \cite{davis07}. Our main results can be summarized as follows:
\begin{enumerate}
\item We showed that quasar color variability, $s_{\lambda_1\lambda_2}\equiv \frac{\partial m_{\lambda_2}}{\partial m_{\lambda_1}}-1$, is best determined by fitting data in the statistically independent magnitude-magnitude space, rather than in color-magnitude space as many studies have done. Unless care is taken to account for the data correlations, the latter approach may lead to spurious or biased estimates of color variability. The $gr$ and $ui$ color variability for the vast majority of the 9093 quasars $s_{gr/ui}<0$, confirming that quasars get bluer when they get brighter.
\item The color variability as measured in $gr$ and $ui$ space exhibits a distinct redshift dependence, which we could clearly attribute to the effect of emission lines exiting/entering the photometric SDSS bands.
From a set of simple models of spectral quasar color variability, 
a model in which the line and continuum vary in phase but with the line amplitude fixed to 10\% or less of that of the continuum,
is able to reproduce the observed redshift trends in the $gr$ (and $ui$) color variability as well as the observed amount of color variability.
\item The fact that we see clearly the impact of the emission-line fluxes on the broad band photometry through the redshift-dependence of the color variability implies that broad-band reverberation mapping should be possible with the data set at hand. 
\item Correcting for the emission lines leaves us with a sample mean (continuum) color variability of $s_{gr} \equiv \frac{dm_r}{dm_g}-1 = -0.17$ and, analogously,  $s_{ui} = -0.46$.
\item We found that the emission line corrected color variability is independent of $L/L_\textrm{Edd}$ and $M_\textrm{BH}$ in both $gr$ and $ui$: there is no correlation between the mass and luminosity of quasars and their color variability.
\item The color variability, however, does depend on the light curve variability properties \citep[described by a power-law structure function as in][]{schmidt10}.
We found that quasars with large variability amplitudes ($A'$) tend to have less color variability, 
as compared to quasars with small variability amplitudes.
\item We found that the characteristic color variability on timescales of years of the individual quasars is larger than the dependence of the typical quasar colors on their overall accretion state (i.e., $L/L_\textrm{Edd}$).
This implies that changes in the overall accretion rate cannot explain the observed color variability. 
Ephemeral hot spots may however be a plausible explanation for the observed color variability.
This picture is confirmed by our comparison of the observed color variability to sequences of steady-state accretion disc models by \cite{davis07} with varying accretion rates, which also exhibit much less color variability as a function of accretion rate.
\end{enumerate}

Our analysis provides a clear indication that on time-scales of years quasar variability does not reflect changes in the mean accretion rate. Some other mechanism must be at work; presumably some disc instability. What mechanism match the existing data, certainly warrants further modeling.
The current study can also be viewed as an initial foray into the realm of multi-band, multi-epoch panoptic photometry that the Pan-STARRS and LSST surveys can bring to full fruition.

\section*{Acknowledgments}
We would like to thank S. W. Davis for providing us with the model output, for the comparison performed in Section~\ref{sec:models}. 
Furthermore we would like to thank Joseph F. Hennawi, Stefan Wagner and Ramesh Narayan for valuable discussions.
KBS is funded by and would like to thank the Marie Curie Initial Training Network ELIXIR, which is 
funded by the Seventh Framework Programme (FP7) of the European Commission. KBS is a member of the International Max Planck Research School for Astronomy and Cosmic Physics at the University of Heidelberg (IMPRS-HD), Germany.
DWH is supported by a Research Fellowship of the Alexander von Humboldt Foundation.

Funding for the SDSS and SDSS-II has been provided by the Alfred P. Sloan
Foundation, the Participating Institutions, the National Science Foundation,
the U.S. Department of Energy, the National Aeronautics and Space
Administration, the Japanese Monbukagakusho, the Max Planck Society, and the
Higher Education Funding Council for England. The SDSS Web Site is
\verb+http://www.sdss.org/+.

The SDSS is managed by the Astrophysical Research Consortium for the
Participating Institutions. The Participating Institutions are the American
Museum of Natural History, Astrophysical Institute Potsdam, University of
Basel, University of Cambridge, Case Western Reserve University, University of
Chicago, Drexel University, Fermilab, the Institute for Advanced Study, the
Japan Participation Group, Johns Hopkins University, the Joint Institute for
Nuclear Astrophysics, the Kavli Institute for Particle Astrophysics and
Cosmology, the Korean Scientist Group, the Chinese Academy of Sciences
(LAMOST), Los Alamos National Laboratory, the Max-Planck-Institute for
Astronomy (MPIA), the Max-Planck-Institute for Astrophysics (MPA), New Mexico
State University, Ohio State University, University of Pittsburgh, University
of Portsmouth, Princeton University, the United States Naval Observatory, and
the University of Washington.
\begin{appendix}
\section{Fitting in Color-Magnitude and Magnitude-Magnitude Space}
\label{sec:fitting}

As mentioned in the text the photometric errors of the $(g-r)$ color are correlated with the photometric errors of the $g$ and $r$ band. Thus, when estimating the color variability of objects in general, and quasars in particular, as is done in the present paper, care has to be taken that the co-variances of the errors are either removed, taken into account or avoided. Above we have avoided the co-variances by estimating the color relation in magnitude-magnitude space and then `translated' that into a color variability in color-magnitude space as described in the text. 
In the following we illustrate the problems one can run into, if the color variability is instead estimated directly in color-magnitude space.

\begin{figure*}
\epsscale{1.10} 
\plottwo{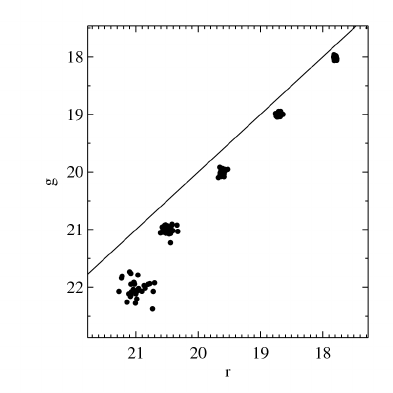}{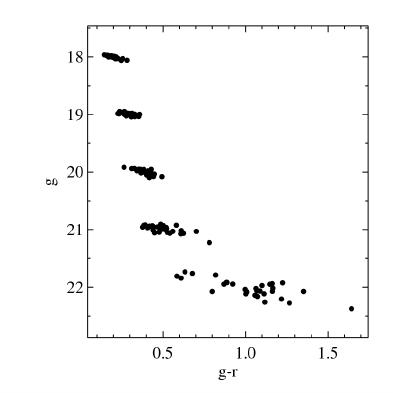}
\caption{Simulated data illustrating the effect of error covariances between magnitude and color, that has not been accounted for in a number of previous analyses. The left panel shows 5 `clouds' of simulated observations in $r$-$g$ space drawn from 2D gaussian distributions at $g=18$, 19, 20, 21 and 22 with `photometric' errors of 0.02, 0.025, 0.04, 0.06 and 0.15 respectively. In the right panel these data are plotted in the $g$-$(g-r)$ color-magnitude space. The stripy pattern seen here is due to the correlation between the errors on the $g$ magnitude and the $g-r$ color. Our test showed that this effect need to be accounted for when analyzing data of SDSS Stripe 82 quality, as is also illustrated in Figure~\ref{fig:fitcom}.}
\label{fig:simdat}
\end{figure*}

The correlated errors between for instance the $g$ band and the $(g-r)$ color are easily illustrated, by simply drawing a set of random `observations' from a gaussian distribution with a standard deviation corresponding to the approximate photometric error at the given magnitude. In the left panel of Figure~\ref{fig:simdat} such a sequence of simulated data is shown. The data have been drawn from 2D gaussian distributions in $g$ and $r$ with mean magnitudes of approximately 18, 19, 20, 21, and 22  and estimated errors of 0.02, 0.025, 0.04, 0.06, and 0.15 respectively. The solid line shows the $g=r$ relation for reference. The color trend (deviation of the data from the $g=r$ line) has been put in to mimic the average color trend of quasars at the given magnitudes. Plotting these simulated observations in color-magnitude space, as done in the right panel of Figure~\ref{fig:simdat}, clearly illustrates the error correlations. The stripy pattern of the color-magnitude diagram is not a consequence of a color change in the object, but a consequence of the fact that the errors in $(g-r)$ are correlated with the errors in $g$; this is the reason that the `length' (or artificial color change) of each set of points grows for fainter magnitudes. 

\begin{figure*}
\epsscale{0.95} 
\plottwo{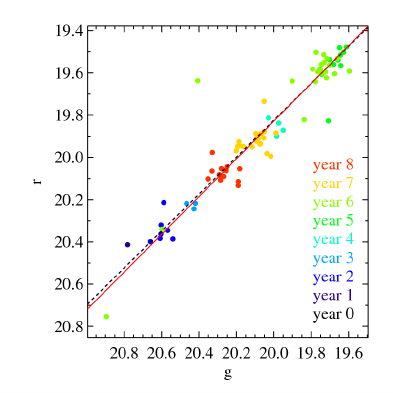}{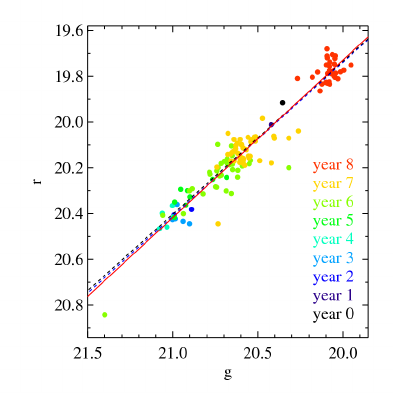}
\plottwo{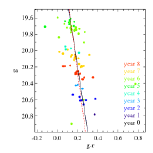}{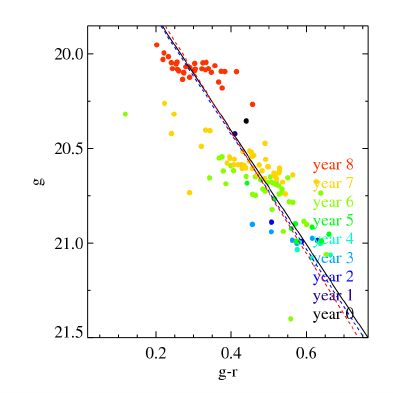}
\plottwo{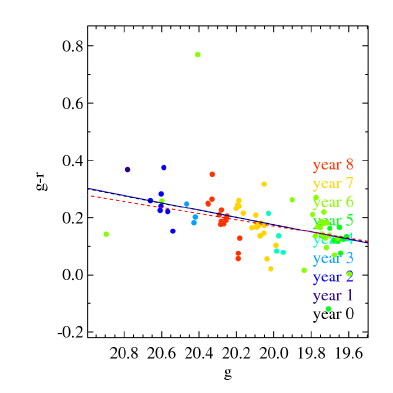}{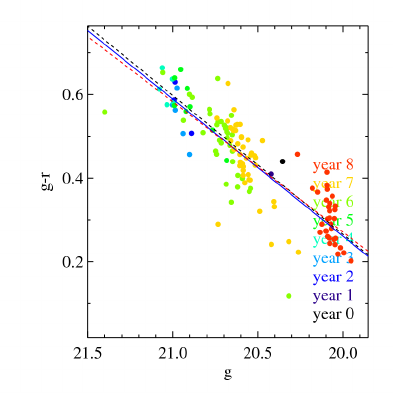}
\caption{Comparison between fitting in magnitude-magnitude, magnitude-color and color-magnitude space. The objects SDSS J0320-0051 (left) and SDSS J2141-0050 (right) from the SDSS Stripe 82 sample are shown in the $r$-$g$ (top), $g$-$(g-r)$ (center) and $(g-r)$-$g$ (bottom) spaces. The observations are color coded according to the observation year. Each plot has one solid and two dashed lines. The solid line corresponds to the fit (performed as described in Section~\ref{sec:fit}) done in the given space, whereas the dashed lines are `translations' from the two other spaces into the present one. The red line has been fit in $r$-$g$ space (top), the black line in $g$-$(g-r)$ (center) and the blue line in $(g-r)$-$g$ (bottom). The stripy-ness in the bottom two panels comes mostly from the correlated errors between $g$-$(g-r)$ as illustrated in Figure~\ref{fig:simdat}. 
}
\label{fig:fitcom}
\end{figure*}

In Figure~\ref{fig:fitcom} we show how this effect looks when dealing with real data. Figure~\ref{fig:fitcom} shows the SDSS S82 photometric data of the quasars SDSS J0320-0051 (left) and SDSS J2141-0050 (right) in magnitude-magnitude, magnitude-color and color-magnitude space. The top panel corresponds to the left panel of Figure~\ref{fig:simdat}. The stripy nature of the data when turned into colors is clearly visible in the center and bottom panels of Figure~\ref{fig:fitcom}. The data have been color coded according to the observation time. In each panel one solid and two dashed lines are shown. The solid line has been fitted to the shown data, whereas the dashed lines are `translations' of the fits from the other two spaces. It is clear that in the case of the superb data set of SDSS S82 the difference between fitting in $r$-$g$ (top), $g$-$(g-r)$ (center) and $(g-r)$-$g$ (bottom) space is negligible. However, if one imagines that only data from year 6 and 7 were available for SDSS J2141-0050 (right panel) and the color variability was estimated based on either $g$-$(g-r)$ or $(g-r)$-$g$, it is clear that the fit would deviate significantly from the `real' color relation because of the co-variant errors.

If this effect is not taken into account or avoided when estimating color variability, and the error co-variance `color change' is interpreted as an actual color change of the quasar, there is a high probability that the results and conclusions will be erroneous.
As mentioned in the text color variability has been estimated in color-magnitude space in the past with only a few exceptions. Hence, there might be cases in the literature where this effect has not been taken properly into account and therefore might affect the validity of the results. It is hard to quantify how much this effect will affect the results and conclusions made so far in the color variability literature, and we have therefore not made any attempts at quantifying it, but will just note that one needs to take this error correlation into account or, as it is done here, estimate the color variability in magnitude-magnitude space to avoid it.

\section{Simplistic Spectral Variability Model}
\label{app:varmod}

In the following we describe the simplistic model for the spectral (color) variability of the quasars used in Section~\ref{sec:varmod}. The simple spectral variability model is an attempt to reproduce the observed redshift trends in the mean color variability, $\langle s_{gr/ui}\rangle (z)$. The model is based on the composite SDSS spectrum from \cite{vandenberk01}, $F_\textrm{VdB}$. We decomposed the composite spectrum in a line component, $F_\textrm{line}$, and a continuum component, $F_\textrm{cont}$, such that
\BE
F_\textrm{VdB} = F_\textrm{line} + F_\textrm{cont} \quad .
\EE
The underlying continuum of the composite spectrum is well modeled (for $1150\textrm{\AA} \lesssim\lambda_\textrm{rest}\lesssim4500\textrm{\AA}$) by a simple power-law with a power-law index of $\beta_\lambda=-1.528$. By fixing the power-law continuum model with a pivot point in the IR (ensuring that $s_{gr/ui}<0$) we simulate the variable continuum by a (time)sequence of power-laws with different $\beta_\lambda$.
Adding fractions of $F_\textrm{line}$ to each power-law simulates the (potentially) variable emission lines. The amount of variability in the emission lines can be fixed to the variability of the continuum power-law via a constant $\alpha$ in Equation~\ref{eqn:alpha}, which as a reminder reads
\BE
\alpha = \frac{\delta F_\textrm{line}}{\delta F_\textrm{cont}} \quad .  \tag{\ref{eqn:alpha}}
\EE
We define
\BEA
\delta F_\textrm{line} &\equiv& \int_{\lambda_\textrm{min}}^{\lambda_\textrm{max}}F_\textrm{line}(\lambda,t_j)\;d\lambda-\int_{\lambda_\textrm{min}}^{\lambda_\textrm{max}}F_\textrm{line}(\lambda,t_{j-1})\;d\lambda \\
\delta F_\textrm{cont} &\equiv& \int_{\lambda_\textrm{min}}^{\lambda_\textrm{max}} F_\textrm{cont}(\lambda,t_j)\;d\lambda- \int_{\lambda_\textrm{min}}^{\lambda_\textrm{max}} F_\textrm{cont}(\lambda,t_{j-1})\; d\lambda 
\EEA
where $\lambda_\textrm{min}\sim 1150\textrm{\AA}$ and $\lambda_\textrm{max}\sim 4500\textrm{\AA}$. The $t_j$ and $t_{j-1}$ refers to the `epochs' of the variability model. This implies from Equation~\ref{eqn:alpha}, that for a fixed $\alpha$ the emission line response for each variability model `epoch' is given by
\BE
\int_{\lambda_\textrm{min}}^{\lambda_\textrm{max}} F_\textrm{line}(\lambda,t_j)\; d\lambda =  
\int_{\lambda_\textrm{min}}^{\lambda_\textrm{max}} F_\textrm{line}(\lambda,t_{j-1})\; d\lambda +
\alpha \int_{\lambda_\textrm{min}}^{\lambda_\textrm{max}} F_\textrm{cont}(\lambda,t_j)-F_\textrm{cont}(\lambda,t_{j-1}) \; d\lambda \quad .
\EE
Here the emission lines are assumed to respond instantly to the continuum variation. By integrating the obtained spectra at the different epochs, $t_j$, over the SDSS bands the model color variability can be estimated. The results shown in Figure~\ref{fig:varmod} are for a variability model with fixed $\alpha$, since such a variability seems to resemble the observed redshift dependence of the color variability the closest.

The pivot point for the results shown in Figure~\ref{fig:varmod} was put at $4\lambda_\textrm{max}\sim1.8\micron$. Changing the pivot point slightly changes the amplitude of the obtained model predictions (dashed and colored curves in Figure~\ref{fig:varmod}) and the mean color variability in such a way that pivot points in the far-IR result in smaller amplitudes and larger $\langle s_{gr/ui}\rangle(z)$. The actual curvature of the curves in Figure~\ref{fig:varmod} does not change with the pivot point, i.e., the redshift dependence is independent of the continuum power-law model pivot point.

This spectral variability model predicts the redshift dependence in $s_{gr}$ and $s_{ui}$ equally well.

A model similar to the one presented here, was used in \cite{richards01} to explain the redshift dependences of the SDSS quasar colors.

\end{appendix}

\end{document}